\documentclass[onecolumn]{aastex631}
\usepackage{lineno}
\begin{document}

\def\msol{\hbox{$\hbox{M}_\odot$}}
\def\lsol{\hbox{$\hbox{L}_\odot$}}
\def\kms{km s$^{-1}$}

\title{Populations of Magnetized Filaments in the\\ 
       Intracluster Medium and the Galactic Center
%\title{Synergies Between Populations of Intracluster Medium\\
%       and Galactic Center Magnetized Filaments
}

\author[0000-0001-8403-8548]{F. Yusef-Zadeh} 
\affiliation{Dept Physics and Astronomy, CIERA, Northwestern University, 2145 Sheridan Road, Evanston , IL 60207, USA
(zadeh@northwestern.edu)}

\author[0000-0001-8403-8548]{R. G. Arendt} 
\affiliation{Code 665, NASA/GSFC, 8800 Greenbelt Road, Greenbelt, MD 20771, USA}
\affiliation{UMBC/CRESST 2 (Richard.G.Arendt@nasa.gov)}

\author[0000-0001-8403-8548]{M. Wardle} 
\affiliation{School of Mathematical and Physical Sciences,  Research Centre for Astronomy, Astrophysics
and Astrophotonics, Macquarie University, Sydney NSW 2109, Australia, (mark.wardle@mq.edu.au)}

\begin{abstract} Magnetized radio filaments are found in abundance in the inner few hundred pc of our Galaxy. Progress in understanding this 
population of filaments has been slow over the last few decades, in part due to a lack of detection elsewhere in the Galaxy or in external 
galaxies. Recent highly sensitive radio continuum observations of radio galaxies in galaxy clusters have revealed remarkable isolated 
filamentary structures in the ICM that are linked to radio jets, tails and lobes. The origin of this class of filaments is not understood 
either. Here, we argue that the underlying physical mechanisms responsible for the creation of the two populations are the same because of their 
similarity in morphology, spacing between the filaments, aspect ratio, magnetic energy densities to thermal pressure of the medium, and that 
both populations have undergone synchrotron aging. These similarities provide an opportunity to investigate the physical processes in the ISM 
and ICM for the first time. We consider that the origin of the filaments in both the GC and ICM is a result of the interaction of a large-scale 
wind with clouds, or the filaments arise through the stretching and collection of field lines by turbulence in weakly magnetized medium. We 
examine these ideas toward four radio galaxy filaments associated with four radio galaxies IC 40B, IC 4496, J1333–3141, ESO137-006 and argue 
that much can be understood in the future by comparing these two populations of filaments. \end{abstract}

%the cooling length to the length of the filaments and streaming transport. 

%We consider 
%that this pressure acts like a wind that can interact with the jet and the lobe and trigger escape of cosmic ray 
%particles that are loosely bound. 
%An alternative scenario 
%that has been proposed for the GC filaments is the collection and draping of field lines by a moving stellar wind 
%source or similar obstacle with respect to the medium.  

%argue that the class of Galactic Center filaments is scaled-down version of radio galaxy filaments in terms of 
%their physical characteristics.

%\keywords{cosmology: observations --- diffuse radiation --- zodiacal dust}

\section{Introduction} 
\label{sec:intro}
In the last few years, sensitive high resolution radio continuum observations of galaxy clusters have begun to 
reveal a population of magnetized filamentary structures in their intra-cluster medium
(ICM) \citep{shimwell16,ramatsoku20,vanweeren21,fanaroff21,condon21,brienza22,giaci22,knowles22,rudnick22}. 
Discoveries of new structures in the ICM surrounding radio galaxies (RGs) are in large part due to
recent advances in broad-band correlators installed on radio telescopes, providing remarkable brightness sensitivity 
and dynamic range with exquisite details at $\sim6''$ spatial resolution. A
striking aspect of the appearance of these long and narrow isolated filaments is that they are not located within 
the radio lobes, the tails of radio galaxies or associated with the radio core of
the Galaxy \citep{lane02,hardcastle19}. The filaments are detected outside radio galaxies but linked to radio lobes 
and jets with a wide range of angles to the orientation of radio jets or the
lobes. The nature of this new phenomenon, which appears to be common in high dynamic range images of radio galaxies, 
had not been appreciated or predicted in previous studies due to lack of sufficient surface brightness sensitivity and angular 
resolution, particularly at lower frequencies.

%An  understanding of filamentary structures in
%RGs is important because they act as excellent probes of the physical conditions of the magneto-ionized medium of 
%the cluster.

Closer to home, in the vastly different environment of the Galactic center (GC), radio observations have uncovered a 
population of magnetized radio filaments with linearly polarized synchrotron emission tracing cosmic ray activity 
throughout the inner few hundred pc of the Galaxy 
\citep{zadeh84,liszt85,bally89,gray91,haynes92,staguhn98,lang99,larosa01,larosa04,zadeh04,nord04,law08,pound18,staguhn19,arendt19}. 
MeerKAT observations have provided a remarkable mosaic of the inner few degrees of the Galactic center revealing 
hundreds of radio filaments, housed within a $\sim400$pc bipolar Radio Bubble filled with thermal X-ray gas that surrounds the Galactic 
center 
\citep{heywood19,heywood22,zadeh22a,zadeh22b,zadeh22c}.  $H_3^+$ measurements toward the Galactic center have 
quantifield the cosmic 
ray ionization rate to be 10$^2$ to $10^3$ times greater than across the rest of the Galaxy
\citep{oka05,oka19}. This high ionization rate 
led to the interpretion that cosmic-rays drive a wind, producing the Radio Bubble \citep{zadeh19}.

%Each of the filaments tracks a source of cosmic rays suggesting high cosmic-ray flux. 
%showing high cosmic-ray flux in the Galactic center.

In spite of decades of studying the GC filaments, a good understanding of these structures is still lacking. 
The emerging  population of extragalactic radio filaments appears to
have remarkably similar properties to the radio filaments in the Galactic center. The comparison of generic 
properties 
of both populations provide an opportunity to gain insight into the origin of both
populations.  Here, side-by-side images of four examples of GC and ICM filaments are displayed in \S2, to illustrate 
their similar morphology, followed by a discussion of the properties that are
common in two populations of GC and ICM filaments in \S3. We then compare the physical  parameters of the two 
populations of filaments in \S4, and discuss their implications for models of their
structure and origin in \S5.

%\vspace{-1cm}

\section{GC versus  ICM  filaments: four examples}

The GC filaments are rich in their morphological variety and details as well as being abundant in the inner few 
hundred pc of the Galactic center. ICM filaments are being revealed in high dynamic images of radio galaxies and 
their surrounds. We have selected a subset of GC filaments based on their similar morphology to ICM filaments. The 
comparison of limited number of ICM filaments with a large number of GC filaments offers an opportunity to advance 
our understanding of the origin of the filaments and to probe the magneto-ionized media of the GC and ICM. We 
examine below the morphology and properties of filaments associated with four RGs and compare them with seven GC 
filaments.

\subsection{3C40B in Abell 194 vs G359.132-0.296 (the Snake), G359.808+0.130 (the Flamingo), and G359.717+0.228 (the
Eyebrow)}

Figure 1 displays a comparison of the filamentary structures associated with the radio jet of 3C40B and three 
morphologically similar groups of filaments  identified in the
Galactic center \citep{zadeh22b}.  3C40B is a radio galaxy at a distance of $\sim75.5$ Mpc, embedded in a poor 
cluster environment Abell 194 \citep{knowles22} which has recently been studied in
detail \citep{rudnick22}. Two remarkable filaments emerge from a deviated region of the jet and run parallel to each 
other as they bend together with two different curvatures leading to  diffuse ends. The
angular spacing between the filaments is $\sim30''$ ($\sim$11 kpc), their widths range between 0.85 and 2.8 kpc and 
both have an extent of $\sim220$ kpc \citep{rudnick22}. The filaments show signs
of physical interaction with a region of the the jet that is highly bent,
suggesting a scenario in which a cloud is crossing the jet and the filaments are then dragged
from the interaction site \citep{rudnick22}.

The Snake has a length of $\sim30'$ ($\sim70$ pc), is one of the longest in the Galactic center, and is 
characterized by two kinks and three different curvatures along its length. Another two
examples of GC filaments are the groups known as the Flamingo and the Eyebrow. The Flamingo shows multiple 
 filaments with similar curvatures and the Eyebrow consists of two
parallel filaments with the same curvature suggesting that they  are associated with  each other.

The morphology of the four filament groups shown in Figure 1 are similar to each other to the extent that it is 
difficult to distinguish which one is in a galaxy cluster based on appearance alone.  The 
systems of filaments in the GC and 3C40B have similar morphologies, surface brightnesses, length-to-width ratios and 
angular spacings.  What is remarkable is that the ICM filaments in 3C40B
are $\sim7000$ times longer than nonthermal filaments (aka NTFs)  in the GC with coherent magnetic structure, 
steepening spectral index, streaming cosmic-ray particles over such a large distance (\cite{rudnick22}) and yet they 
show similar physical characteristics.

The ranges of the spectral indices are $\alpha$ = [-0.60,-0.79] for the Snake, $\alpha$ = [-0.80,-1.05] for the 
Flamingo, $\alpha$ = [-0.51,-0.76] for the Eyebrow \citep{zadeh22b}. These estimates
of spectral index are made along individual filaments. The GC spectral index values are all flatter  than 
the spectral indices of the filaments  in 3C40B which have  a mean value of
$\alpha\sim-2$ and steepen from -1.3 to -2.5
with  distance from the jet. 

%The  variation of the spectral index of the Snake as a function of Galactic latitudes is noted between 
%[-0.76,-0.51] \citep{zadeh22b}.

The equipartition magnetic field of GC filaments
ranges  between 20-100 $\mu$G (for a cosmic ray proton to electron 
ratio of $p/e=1$) which is much higher than that of 3C40B with the magnetic field of 2-7$\mu$G. The equipartition 
magnetic field of the Snake has a maximum of $\sim0.1$ mG and decreases away from the Galactic plane 
\citep{zadeh22a}. Furthermore, the direction of the magnetic field runs along the 3C40B and the Snake filaments.  
(The magnetic field directions of the Flamingo and Eyebrow are unknown).

\vfill\eject

\subsection{Radio galaxy IC 4296 vs G359.992-0.572 (The Comet Tail)}

IC 4296 is a low-luminosity elliptical galaxy, in  the cluster A3566 at a distance of 50 Mpc
\citep{condon21}. A sensitive radio image of this galaxy shows a remarkable group of three filaments with angular
spacing of $\sim20''$ (called threads in \cite{condon21}) that appear to converge to a point.  The projected lengths
and widths of the filaments are $\sim50$ and $\sim2$ kpc, respectively, with surface brightness of $\sim0.1$ mJy
per $\sim7.6''$ beam \citep{condon21}. This group of filaments lies to the south of the 
western
radio jet where the jet shows a wiggle, thought to be due to Kelvin-Helmholtz (KH)  instability \citep{condon21}.
These authors suggest that  relativistic electrons escape from the jet where there is KH 
instability
along the radio jets \citep{condon21}. There is also a single isolated filament with different position angle that
arises from the eastern jet. One puzzle is that the spectral index of the filaments steepens  from 
$\alpha\sim -1$ to -1.5  closer to the injection point of the radio jet.

The GC grouping of the Comet Tail consists of six linear structures which resemble a fragmented comet tail. 
Some of the filaments branch out into fainter components. The spectral 
index of
individual filaments vary and show steep spectra in the range [-1,-1.8].  There is  a trend that the
GC filaments with steeper spectral indices lie at high latitudes \citep{zadeh22a}.
The comparison indicates similar morphology and steep spectral indices in  both the Comet Tail and IC
4296 filaments.

%The spacing between the Comet Tail filaments is relatively large compared to the mean angular spacing of $\sim17''$
%typically found for GC  groupings. 

\subsection{The core of A3562 galaxy cluster vs  G359.411-0.709 (The Feather)}

Figure 3a shows a radio image of the core of the A3562 galaxy cluster at a distance of $\sim200$ Mpc 
\citep{giaci22}. A linearly polarized  filament with a
length of 50 kpc   runs perpendicular to the end of the tail of the radio galaxy J1333-3141. The 
filament that arises from the  region where  the jet terminates is distorted before it  splits into two
components \citep{giaci22}.  The splitting occurs 60 kpc away from the radio tail. The mean spectral index of the 
filament is $\alpha\sim-1.5$.

For comparison we show in Figure 3b, a striking GC filament, the Feather which splits into a two-pronged forked 
filament with the junction at the location of a compact source G359.416-0.706 with possible stellar 
counterpart detected at IR \citep{zadeh22b,zadeh22c}. The compact source 
is suggested to be an obstacle that splits the bright filament into two parallel, fainter components with steeper 
spectral index values. The comparison noted in Figure 3 provides another example that radio filaments in both 
classes of Galactic and extragalactic filaments break up into multiple components, suggestive of a flow along the 
filaments that splits into two components as it runs into an obstacle.

\subsection{ESO137-006 vs G359.484+0.1122 (The Bent Harp)}

Figure 4a shows a remarkable group of filaments linked to the radio lobes of ESO137-006, a luminous radio galaxy in 
the Norma galaxy cluster at  70 Mpc from us \citep{ramatsoku20}. The wide-angle
tail morphology of this radio galaxy includes  multiple straight and bent filaments extending from the eastern lobe 
toward the western lobe. The longest straight filaments act as a bridge connecting
the two radio lobes. In addition, a number of semi-regularly-spaced filaments are noted to the south of the lobes. 
The length and width of the longest filament are  80 and 1.2 kpc, respectively, with
an aspect ratio (length:width) of 67 \citep{ramatsoku20}. The mean spectral index of the filaments, 
$\alpha\simeq-2$,  is similar to the spectral indices of the lobe from which they emerge.
The western half of the galaxy is completely surrounded by diffuse X-ray emission. The western lobe is also highly 
distorted where diffuse and filamentary structures are present. The filaments
bridging the lobes, with similar morphology to the above three examples, arise from a region that the jet is 
expanding into the lobe.

G359.484+0.1122 consists of five regularly spaced filaments with the appearance of a bent harp 
\citep{thomas20,zadeh22b}. A trend was noted between the length of the filaments and their brightness
as well as with the steepening of the spectral index in the range between [-1.3,-1.0] \citep{zadeh22b}.

The longest and straightest filament in ESO137-006 (bridging the radio lobes) appears like many straight GC filaments running 
perpendicular to the Galactic plane \citep{zadeh22b}.  
The Bent Harp is one group of GC filaments with a resemblance to the ICM structures  in ESO137-006.   
Unlike the filaments of ESO137+006, the Bent Harp is not connected to any obvious sources of cosmic rays.

%The comparison of three filaments called collimated synchrotron threads (CST) CST1, CST2 and CST3 in 
%\cite{ramatsoku20}, can be made with the Bent Harp.

\vfill\eject

\section{General physical properties}

Galactic and extra-galactic filaments are produced in totally different environments, namely, the nucleus of a 
normal galaxy and the intracluster medium.  Their length scales, energy densities, and other characteristic 
parameters differ by orders of magnitude. In spite of these differences, we argue that similar processes operate in 
both systems motivated by their similar morphology as well as similar dimensionless ratios of their physical 
parameters. We will examine and compare four examples of GC and RGs in the next section, following the common 
physical characteristics, as described below.

$\bullet$ There is a remarkable similarity in morphological details of the GC and ICM filaments despite their 
distances differing by a factor of
 $\sim10^{3-4}$. Both classes of filaments are narrow
and long with typical aspect ratios ranging between 10 to 50, assuming that the widths of the filaments are 
resolved. The filaments show
bending, and wiggles along their lengths and have similar
surface brightness roughly ranging between 0.01 and 0.1 mJy beam$^{-1}$. 

$\bullet$ Both classes of filaments appear either as single isolated filaments or multiple filaments grouped 
together. Some of the apparently single filaments of the GC 
are found to consist of bundles of filaments in higher 
$\sim1''$ resolution VLA observations.  
Many filaments in groupings are approximately equally separated from each other and run 
parallel to 
each other, giving a harp-like appearance. Filaments in some groupings converge to a point or to an extended source. 
Groups of filaments shift sideways together, changing direction coherently, implying that they are parts of the same 
system of filaments with similar origin.

$\bullet$ The mean spacings between parallel filaments in a given GC grouping peaks at $\sim 16''$ \citep{zadeh22b}. 
There is  
a limited number of groups of ICM filaments but they show regular spacing between parallel
filaments, as noted in numerous groups of GC filaments. In the case of 3C40B, the spacing between two parallel 
filaments is $\sim 30''$ \citep{rudnick22}. We expect that the spacing to length ratio
of ICM and GC filament to be similar to each other if similar processes operate in both systems.

$\bullet$ Both systems of GC and ICM filaments trace synchrotron emission from cosmic-ray electrons.  They both show 
a high percentage of polarization, $\sim50\%$, revealing ordered magnetic fields
(B) running along the filaments. The orientation of the field is expected from velocity shear at the interface 
between the ICM filaments and the ambient gas in the ICM \citep{condon21}.

$\bullet$ The mean magnetic field strengths along the GC filaments range between $\sim100$ to 400 $\mu$G depending 
on the assumed ratio of cosmic-ray protons to electrons (p/e). The magnetic field
of ICM filaments is typically estimated to be a few $\mu$G, 50 to 100 times smaller. The low magnetic field of ICM 
filaments is expected because 
the filaments have  large depth L (see equation 1), and  old if steep spectral index, where flux density
$S_\nu\, \propto\, \nu^{\alpha}$, is an indication of age of cosmic-ray electrons. The weaker magnetic field 
in the ICM filaments imply long cooling time scales which scales as $B^{-3/2}$ for observations at a fixed frequency. 

%of cosmic-ray electrons, which depends inversely on the strength of the magnetic field ($\tau\propto B^{-3/2}$).

$\bullet$ The mean spectral indices of the GC filaments are steeper than supernova remnants (SNRs, 
$\alpha\sim-0.62$)  with a value of $\alpha\sim-0.8$. Furthermore, a trend is noted in steepening
of the spectral index as a function of absolute Galactic latitude but not necessarily for typical GC filaments \citep{zadeh22a,larosa00}. 
With limited number of ICM filaments, the typical  spectral indices  appear steeper 
than GC  filaments, which is not surprising given the vastly longer time scales. 
In addition, the GC filaments show a number of 
filaments, 
such as the radio Arc near l$\sim0.2^\circ$, with 
relatively flat spectral  index \citep{heywood22,zadeh22a}. 

%which could be explained by older ages of ICM 
%filaments relative to their synchrotron loss time scales.

$\bullet$ In some of ICM filaments, radio lobes and tails of radio galaxies appear to provide injection of cosmic 
ray particles to illuminate radio filaments. However, the sources of particle
injection in the GC filaments has not been well established. Recent MeerKAT measurements indicate a number of 
compact radio sources located at one or other end of the GC filaments. This suggests that compact
radio sources are potentially responsible for injection of cosmic ray particles into radio filaments or acting as 
interfaces to collimate cosmic-ray driven winds.

$\bullet$ Diffuse thermal X-ray gas and diffuse nonthermal radio emission in the ISM of the GC and ICM of RGs 
suggest another similarity in the 
physical conditions of these environment.  In addition, there is evidence of 
 high cosmic-ray ionization rate from $H_3^+$ studies of  the Galactic center region. 

%Both the GC and ICM environments  are filled with a mixture of hot X-ray gas and diffuse cosmic-ray particles.

%, which could generate large-scale winds.

\vfill\eject
\section{Origin of two populations of filaments}

The previous sections described the overall physical properties of a subset of ICM and GC filaments, as members  of 
two 
populations of magnetized radio filaments. There are hundreds of  radio filaments found
in the Galactic center at similar distance. A statistical study characterizing the mean properties of the spectral 
index, the equipartition magnetic field, the spacing between the filaments and
surface brightness, is  based on MeerKAT data at 1.28 GHz 
\citep{zadeh22a,zadeh22b}. The physical parameters of GC and ICM filaments 
are listed in columns 2 and 3, 
respectively, of Table 1. 
In this table, the range and mean values  of the length, 
width, magnetic field strength, surface brightness and spacing between the filaments are given.  
 The lengths of the GC filaments follow a power-law distribution 
(Yusef-Zadeh et al. 2022, in preparation). 
 At present only a handful of ICM filaments are known, thus there are not sufficient measurements to do statistics 
beyond simple means. This is because ICM filaments require high dynamic range images to uncover faint magnetized 
filaments and only limited observations are presently available. The GC filaments have a wide range of surface 
brightnesses with many more fainter filaments. The faint end is cutoff by limited sensitivity. 
We recognize  that the ICM 
filaments detected so far quite likely are the most prominent members of a larger population, with many fainter filaments 
yet to be detected.

%For GC and ICM filaments that have similar surface brightnesses, the ratio of radio luminosity of 
%ICM to GC filaments is similar to the square of the ratio of their distances, roughly $\sim10^{8-10}$. If the same 
%process operates in producing both GC and ICM filaments, there should be a lot more fainter filaments in the ICM.

%\vspace{-0.5cm}

\begin{deluxetable}{llll}
\tabletypesize{\scriptsize}
\tablewidth{0pt}
\tablecaption{The physical characteristics of the Galactic Center and radio galaxy filaments 
}
\tablehead{
\\
%\cline{2-3}
\colhead{Physical Parameters} &
\colhead{Galactic Center Filaments} &
\colhead{Radio Galaxy Filaments} 
}
\startdata
  length (pc)  & [4,60] &   10$^{3-5}$ \\
  typical width (pc)&     0.5  & few$\times10^{3}$ \\
  typical aspect ratio &    [10,100]  & [10,70] \\
  magnetic field strength (mG) &  [0.1,0.6] &   few$\times10^{-3}$\\
  spectral index ($\alpha$) & $\sim-0.8$ (mean) [-2,0]  &  [-2,-1]\\
  surface brightness (mJy beam$^{-1}$) &  [0.01-10] & [0.01,10]\\
  spacing between filaments (pc) & $\sim$0.7 mean [0.4,1.2] & [10-20]$\times10^3$\\
\enddata
\end{deluxetable}

\label{tab:table1}

\subsection{Physical parameters of GC and ICM filaments}

We compare the measurements and estimates of the physical characteristics of the two populations of filaments at the 
GC and in the ICM of galaxy clusters, as tabulated in Table 1, to explore the
implications for the formation of the magnetic structures and the source of the ultra-relativistic electrons 
responsible for the observed radio emission.  While the filament length scales and
magnetic field strengths differ by orders of magnitude, it is their respective dimensionless ratios of time scales 
and length scales that should be compared. 
This allows us to examine whether the
underlying physical mechanisms responsible for the creation of the two populations are similar.  
In particular we 
are interested in the mechanisms responsible for creating magnetized filaments, and 
for accelerating the relativistic electrons.

First, the morphological parameters of the two populations of filament are similar:  the aspect ratios are $\sim 
30:1$, multiple parallel filaments, splitting into two pronged fork, bending and spacing between 
the filaments. The filaments appear in parallel bundles or converging toward a point 
in both populations with a spacing of a few 
filament widths.  
These similarities are hardly surprising given that the motivation behind making
this comparison is that structures show a remarkable resemblance in the two environments, as shown in Figures 1-4.

Second, the filaments' magnetic field are estimated to be of order 100 and $3\,\mu$G in the GC and ICM, 
respectively.  
Their corresponding magnetic energy densities, $B^2/8\pi \approx 3\times10^6$ and
$3\times10^3$\,cm$^{-3}$\,K,  are comparable with or lower than the thermal pressure of the medium in 
which they are embedded. These characteristics suggest that the filaments are
confined by the thermal pressure of the external medium.  This is supported in the case of the filaments in Abell 
194 and IC 4296 (\cite{rudnick22,condon21}), where the X-ray surface brightness of
the intracluster medium is reduced along the filament, consistent with exclusion of hot gas from the filament's 
interior. In the case of GC filaments, high external cosmic ray energy density inferred from H$_3^+$
measurements contributes to confinement of  the GC filaments \citep{zadeh19}.

Third, the spectral index values are steeper for ICM  filaments than the mean spectral index of the GC filaments, as 
given in Table 1. Both systems of filaments also show steepening of the spectral index along individual filaments.  
The relativistic electron populations responsible for the
observed synchrotron emission  at 1.2\,GHz, have steep spectral indices and show 
high percentage of linear polarization based on sensitive observations for both the GC and ICM filament 
populations. 

For the equipartition field strengths mentioned above, the characteristic energy of the electrons dominating the 
emission at 1.2\,GHz is $E = (4\pi m_e c\nu/3eB)^{1/2}\,m_e c^2 \approx 0.9$ and 5\,GeV, respectively.  The synchrotron 
power radiated by an ultra-relativistic electron of energy $E=\gamma m_e c^2$ is $\dot{E} = 2 \sigma_T c \gamma^2 
B^2/4\pi $ where for simplicity we have assumed a $90^\circ$ pitch angle.  Then the synchrotron loss time scale  $t_s = 
E/\dot{E} \approx 0.9$ or 200\,Myr for the GC or IC filaments, respectively.   As electron acceleration mechanisms tend 
to produce electrons with $E^{-2}$ energy spectrum and hence a $\nu^{-0.5}$ synchrotron spectrum, the steepness of the 
observed spectra imply that the filament ages are longer than the synchrotron life times given above.  In addition the 
ages cannot be too much more than this because the filaments would rapidly become faint.  Thus the synchrotron life 
times give a 
rough idea of the age of the filaments, and impose a requirement on the cosmic-ray electron transport mechanism which 
must be able to distribute relativistic electrons along the observed filament lengths on this time scale.

\subsection{Transport of cosmic-ray electrons: diffusion versus streaming}

The disparate synchrotron loss time scales need to be placed in context.  For source models involving acceleration 
or injection of relativistic electrons at a localized site, they impose a requirement to be able to transport the 
electrons via diffusion or streaming from the injection site throughout the observed filament lengths $L\sim 30\,$pc 
and $100\,$kpc for the GC and ICM populations, respectively.

If transport is by diffusion, then the diffusion time scale $L^2/\kappa \la t_s$ so the diffusion coefficient 
$\kappa\ga L^2 t_s \sim 3\times10^{26}$ and $2\times 10^{31} $ \,cm$^2$\,s$^{-1}$,
respectively, implying mean free paths $l = 3c/\kappa\sim 0.01$ and 570\,pc.  The required diffusion is plausible 
for GC filaments with  $L \la 30$ pc but seems rather large in the case of the ICM
filaments \citep{rudnick22} as well as GC filaments,  such as the Snake,  with lengths $L \ga 30$ pc.  
(see Fig. 1).

The alternative to diffusion is that the cosmic ray pressure is sufficient to allow them to stream along the 
magnetic field lines, in which case scattering by self-generated waves limits the
streaming to the Alfv\'en speed, $v_A$.  The requirement to populate the entire filament length then implies that 
$L/v_A \la t_s$, i.e. $v_A\ga L/t_s \sim 30$ and $\sim 600$\,\kms\, for the GC and
ICM filaments, respectively.  The density of the ionized thermal plasma present within  the filaments is unknown, 
but we 
can usefully estimate the maximum density able to yield the required Alfv\'en
speeds, using the characteristic field magnetic strengths $B\sim 100$ and 3\,$\mu$G for the two populations.  This 
implies that the thermal electron density $n_e\la 3\times 10^{11}$ and $8\times
10^5 $\,cm$^{-3}$ for the GC and ICM filaments, respectively.  These upper limits clearly exceed any plausible value 
in each of the environments by orders of magnitude, so we can conclude another similarity in parameter space between 
the two populations  that
streaming would be able to transport electrons from their injection site along the entire length of the observed 
filaments. 

\subsection{ICM and GC filament models}

Models of ICM filaments benefit from the fact that there is an obvious reservoir of cosmic-ray particles in jets, 
lobes, mini radio halos \citep{brunetti14}, that can be injected into filaments. For example, the filaments of radio 
galaxy IC 4296 are thought to originate where helical KH instability disrupts the flow of the radio jet followed by 
the relativistic particles feeding the filaments \citep{condon21}. One puzzle is why the spectrum flattens instead 
of steepening moving away from the presumed injection point at the end of the cometary structure's tail towards the 
convergence point of the filaments.  One possible explanation for this is that the injected particles' pitch angle 
would tend to increase to conserve the magnetic moment $p_\perp/B$ as they spiral along the converging field lines.  
This mechanism relies on the initial pitch angle of the injected particles being anti-correlated with energy, and 
pitch-angle scattering to be ineffective in isotropizing the pitch-angle distribution. 

%The change in harder energy spectrum of particles resulting from an anisotropy in pitch angle distribution has been 
%estimated in magnetically dominated plasma turbulence \citep{comisso20}.

In the case of the A3562 galaxy cluster, it has been suggested that the tail of the radio galaxy is interacting with 
a tangential wind blown in the direction along the filaments \citep{giaci22}. An alternative to this picture is a 
scenario in which the cosmic-ray electrons injected from the tail, feed cluster magnetic field lines 
\citep{giaci22}. In another model, cosmic-ray electrons associated with two parallel filaments of 3C40B are injected 
from a region where the jet is highly distorted \citep{rudnick22}. In particular, an impact between a moving dense 
cloud and the jet is expected to distort the shape of the jet as the magnetic field lines are dragged away from the 
site of the interacting jet \citep{rudnick22}. 
Another remarkable object  is 
source C in A2256  with its extremely narrow and long tail \citep{owen14}.  
This unusual source is 
considered to be a one-sided radio jet. However, it has a very similar morphology to 
GC radio filaments. 
Lastly, the collimated synchrotron filaments in ESO 137-006 could be 
due to the interaction of the magnetic fields of the radio lobes with the 
magneto-ionic plasma of the intracluster medium 
\citep{ramatsoku20}.

\subsection{Scenarios for the origin of the ICM and GC filaments}

In GC and ICM populations, there are two broad scenarios that could potentially explain how magnetic filaments are 
formed. Both scenarios posit amplification and organization of an initially weak magnetic field in the surrounding 
medium.  In the first scenario, the filaments are formed by an initially weak field that is amplified and structured by 
subsonic turbulent motions in a weakly magnetized medium, a scenario that has been proposed both for GC 
(\cite{boldyrev06}) and the ICM environments \citep{porter15,vazza18,rudnick22}. The challenge in this picture is that 
simulations show the overall orientation of most filaments is random \citep{porter15,vazza18}. The GC filaments are 
mainly directed in the direction away from the Galactic plane. It is possible that alignment of GC filaments is 
generated by cosmic-ray driven wind running away from the Galactic plane, thus creating a velocity shear in the 
turbulent medium that aligns the filaments.

In the second scenario the filaments are created by subsonic flow of a weakly magnetized medium past an obstacle.  
The passing flow drapes the obstacle with field lines that are then compressed and
stretched back into comet-like tail, creating a magnetic filament with magnetic pressure of order the thermal ram 
pressure in the medium.  The obstacle could be any relatively dense ionized clump
that creates a bow shock in the flow such as a stellar wind bubble, a red giant, an expanding HII region, 
wind tubes associated with red giants in AGN relativistic bubbles \citep{chugai11}, 
or a 
disturbed region along jets or lobes showing size scales similar to the separation between parallel filaments.
In this picture, the magnetized filaments are produced at the interaction sites of a large-scale 
driven wind outflowing away from the Galactic plane and embedded stellar wind
bubbles \citep{rosner96,shore99,zadeh19}. The large-scale Radio Bubble filled with X-ray gas in the Galactic center 
is considered to be arising from cosmic-ray driven outflow \citep{zadeh19}. In
this picture, the obstacle sets the length scale of the separation between the filaments. This is analogous to the 
interaction of fast-moving mass-losing stars whose winds interact with the ISM and
create cometary tails (Martin et al. 2007).
In addition to cometary model, numerous other 
models have been proposed in the past to explain the origin of the GC filaments and 
the sources in 
which cosmic-ray electrons are accelerated. Unlike ICM filaments, it is not clear what feeds cosmic-ray electrons into 
filaments 
\citep{nicholls95,rosner96,shore99,bicknell01,dahlburg02,zadeh03,boldyrev06,ferriere09,barragan16,zadeh19,thomas20,sofue20,coughlin21}.

In the context of the ICM filaments, the relative motion with the external medium is
created by the orbital motion of the obstacle or  a head-tail radio galaxy through the 
ICM around the center of the cluster. 
Additional contribution for slow-moving galaxies close to the center of the 
cluster could be moving through denser ICM \citep{toniazzo01,douglass08}.
Other sources of external pressure could be due
"weather" motion within the ICM \citep{brunetti14}.
On the other hand, in the Galactic 
center ISM, there is contribution due to a 
large-scale nuclear wind generated by high cosmic-ray pressure \citep{zadeh19}.

One interesting aspect of this picture when applied to ICM filaments is that the filaments are linked to jets 
or lobes of radio galaxies. The locations where the filaments cross radio
galaxies are highly distorted and cosmic-ray particles are likely to escape from the jet \citep{condon21}. This 
supports the picture in which there is a concentration of cosmic-ray particles in the
jet or the lobe that injects relativistic plasma into the ICM filaments.  As described in \S4.3, the origin of 
filamentation is interaction with an obstacle and splitting of the filament into
multiple filaments. Another possibility for the origin of the filamentation is synchrotron cooling instability 
formed as a result of the interaction of cosmic-ray generated by the winds \cite[e.g.,][]{simon1967}. 
The mean
spacing and the mean width of the filaments are expected to be similar to each other \citep{zadeh22a}.
In the case of ICM filaments, the spacing is estimated to be the product of the cooling time scale and the Alfv\'en 
speed which is of the order of kpc.

Assuming that the GC and ICM filaments are formed with the same mechanism in the context of an origin involving an 
obstacle collimating the magnetized wind, the lack of compact radio sources
associated with some of the GC filaments is likely to be related to the lifetime  of the obstacle. 
In particular, if an obstacle is an expanding HII region
or a planetary nebula, then it  may have already dissipated as simulations of cloud-wind interactions indicate 
\citep{barragan16}. Another  possibility is 
that the source of injection 
has a very steep spectrum and would  be visible only at very low frequencies.

The above  cometary and turbulent models  are technically different than a scenario in which 
 there is a strong pre-existing organized field 
dominating the region, with  only some field lines being  lit up by an injected
population of cosmic ray electrons.  This model was one of the first models proposed for the GC which was originally 
thought to be threaded by a milligauss  poloidal field \citep{morris07}.  However,
the diffuse radio emission from inner few hundred pc of the galaxy suggests a much weaker global field, $\sim 
10-20\,\mu$G \citep{zadeh13}. This scenario is also inapplicable 
to  the ICM environment in which 
the field strengths  are much less than 
several $\mu$G inferred for the magnetized filaments.

\section{Summary}

%We compared two populations of magnetized filaments which differ vastly in their physical properties, such as 
%lengths, widths and magnetic field strengths. However, we argued that they are analogous to each other because it 
%is 
%dimensionless ratios of physical time scales such as aspect ratio and age-to-cooling time scale ratio that are 
%important. We discussed these ratios, based on self-similar solution applied to physical properties. We then argued 
%that models that have been proposed in one class of filament can be applied to the other.  Lastly, we outlined the 
%existing models for the two filament populations and the synergies between them.

We compared the two populations of magnetized filaments in the Galactic Centre and ICM.  They differ vastly in their 
physical properties, such as lengths, widths and magnetic field strengths. However, we argued that nevertheless, 
they are analogous to each other as might be anticipated based on their similar morphologies.

In both cases, the filaments are in rough pressure equilibrium with their surroundings, but are more strongly 
magnetized, consistent with scenarios in which they are formed by dynamical processes at work in their surroundings. 
In one scenario, which has previously been discussed in both the GC and ICM contexts, the filaments arise through 
the stretching and collection of field lines by turbulence in weakly magnetized medium.  An alternative scenario 
that has been proposed for the GC filaments is the collection and draping of field lines by a moving stellar wind 
source or similar obstacle with respect to the medium.  It is not clear that this mechanism can create the large 
aspect ratios of the ICM filaments, because any orbital motion is mildly transonic at best and so is unlikely to 
lead to a tight magnetized tail or lead to sufficient particle acceleration to energize it.

The source of the relativistic particles still remains to be established.  In the GC, a few filaments have 
associated compact radio sources, and in the ICM the filaments are associated with radio jets and lobes, but the 
exact nature of any connection in both cases remains unclear.  Both filament populations have steep synchrotron 
spectra indicating an aged cosmic-ray electron population, so it is possible that any injection event happened long 
ago and the link to the injecting source is no longer apparent.

The striking similarities between these enigmatic populations, despite their very different environments,  suggests the 
possibility that one or the other may be amenable to observational probes that shed light in the physical processes 
at work in both populations.  Some examples include: (i) more sensitive MeerKAT observations of ICM sources to look 
for 
fainter as well as  shorter (young and bright)  filaments, (ii) high resolution $1''$ VLA study of ICM and GC 
filaments, (iii) X-ray studies searching for 
depressions in X-ray surface brightness overlapping ICM and GC filaments.

\section*{Acknowledgments}
Work by R.G.A. was supported by NASA under award number 80GSFC21M0002. 
We are grateful to   W. Cotton, S. Giacintucci, M. Ramatsoku and L. Rudnick  for providing their  FITS 
images of  their published data.  We thank L. Rudnick and the referee for useful comments. 
The MeerKAT
telescope is operated by the South African Radio Astronomy Observatory, which is a
facility of the National Research Foundation, an agency of the Department of Science and
Innovation.
The National Radio Astronomy Observatory is a facility of the National Science Foundation operated under cooperative 
agreement by Associated
Universities, Inc.

%ApJL\bibitem[\protect\citeauthoryear{Bicknell \& Li}{2001}]{bicknell01} Bicknell G.~V., Li J., 2001, ApJ, 548, L69

\bibliographystyle{mnras}

\vfill\eject

\begin{figure}
\center
\includegraphics[width=7in]{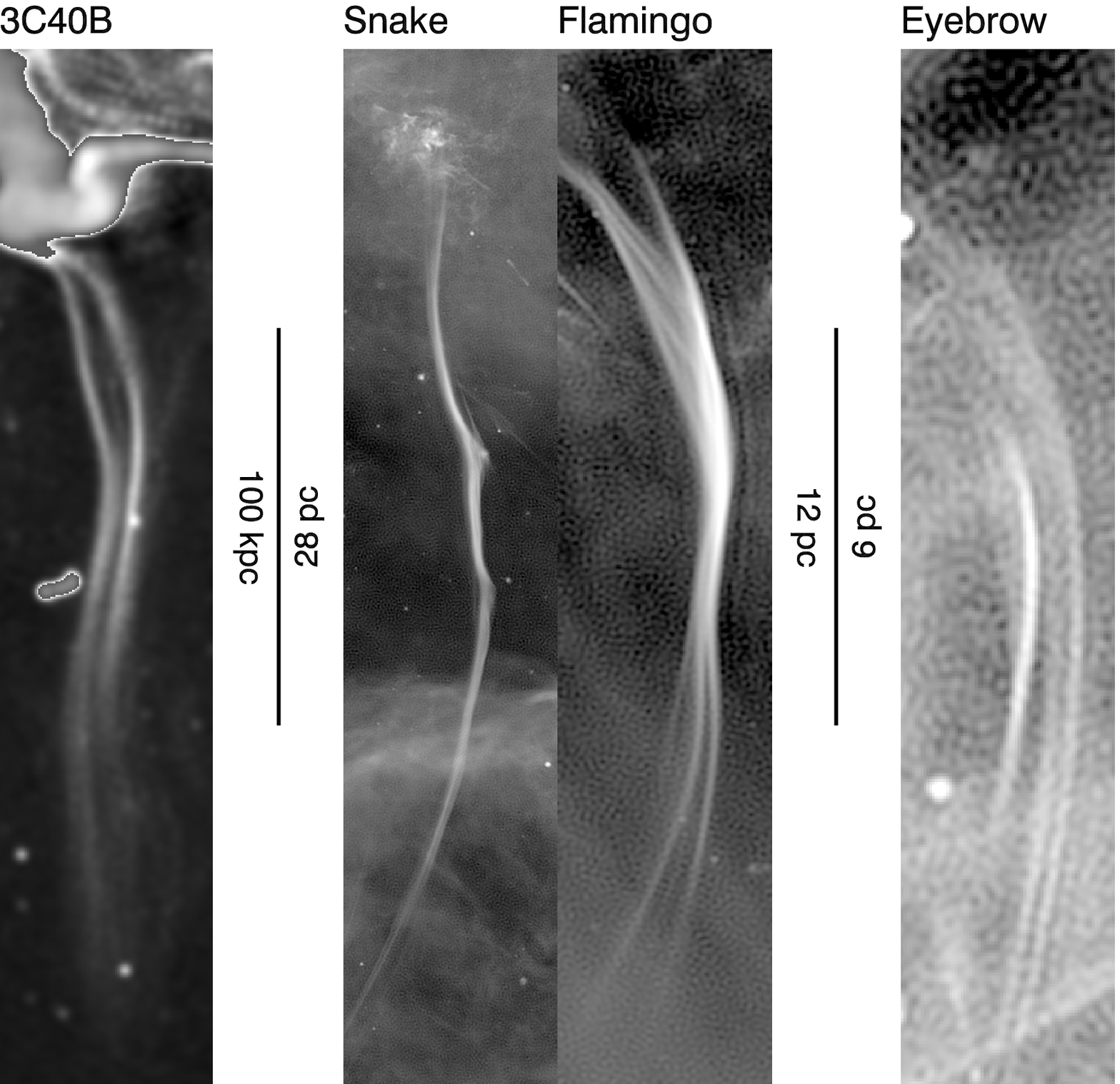}
\caption{
{\it Left}
A grayscale MeerKAT image of the  radio galaxy 3C40B
and its long filaments   \citep{rudnick22}.  
The filaments are connected to a portion of 
the jet shown at the top of the panel. 
{\it Right}
The three panels to the right display three groups of GC radio filaments based on 4$''$ resolution MeerKAT mosaic of 
the   
Galactic center at 1.28 GHz \citep{heywood22},  discussed in detail in  \citep{zadeh22b}.
The intensity I (Jy beam$^{-1}$) scales are:
3C40B $-4 < \log(I) < -1.4$ only for the fainter emission; Snake $-4 < \log(I+3\times10^{-4}) < -2.33$;
Flamingo $(-4.3 < \log(I+3\times10^{-4}) < -2.3$;
Eyebrow  $(-4.3 < \log(I+5\times10^{-4}) < -3.$
}
\label{f:fig1}
\end{figure}

%\begin{figure}[ht!]
\begin{figure}
\center
 \includegraphics[width=6.5in]{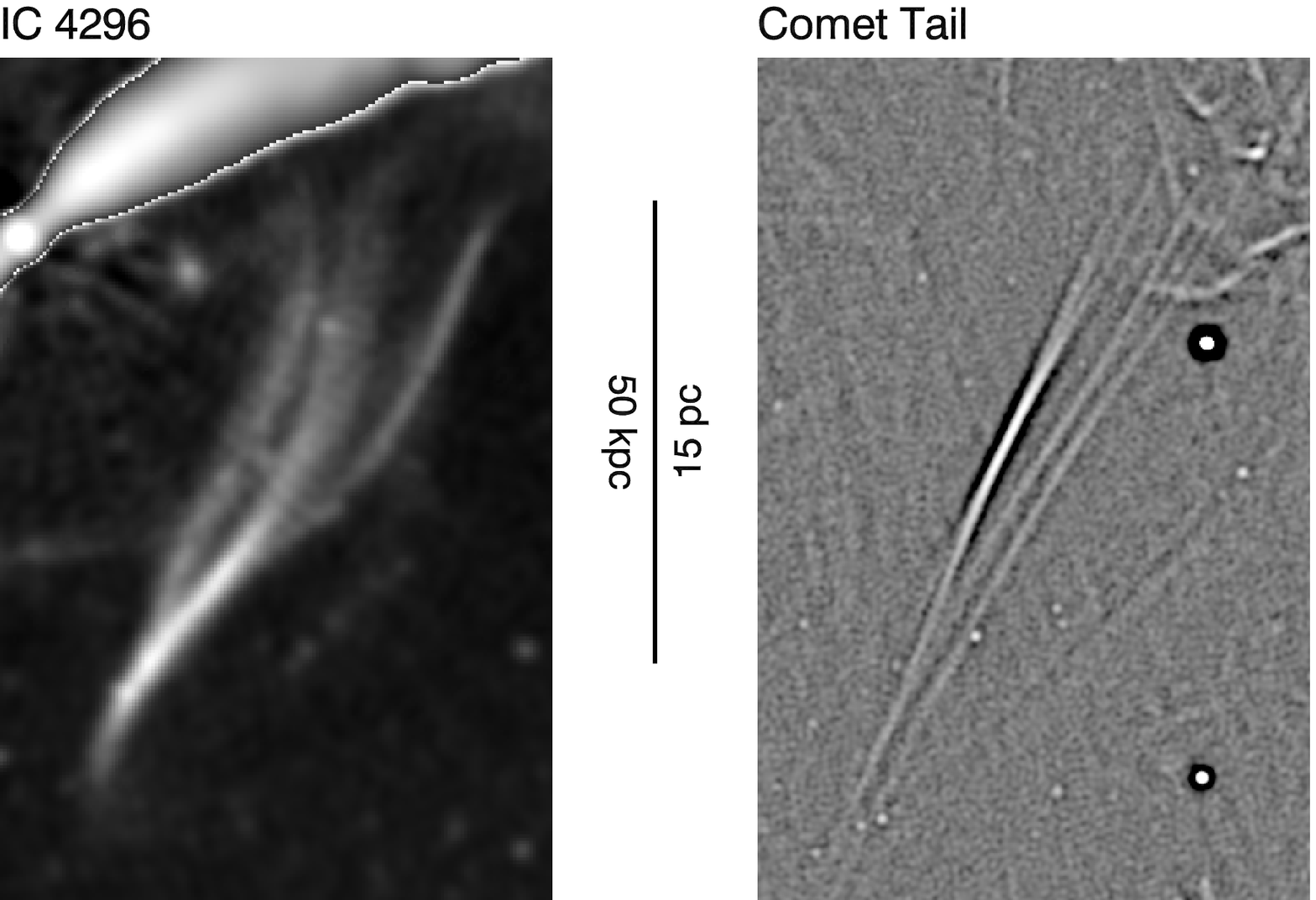}
 \caption{
{\it Left}
A grayscale MeerKAT image of the core and its  radio jet (top of the panel)  and the filaments (center of the 
panel) of IC 4296, an elliptical radio
galaxy, are displayed
 \citep{condon21}. Because of the strong surface brightness of the core and the jet compared to the filaments, 
different color tables were applied.
{\it Right} 
A  MeerKAT filtered image of the Comet tail as discussed in \citep{zadeh22b}.
The intensity I (Jy beam$^{-1}$) scales are:
IC4296 $-4 < \log(I) < -1.2 (I>3.6\times10^{-4})$;
the Comet tail: $-5   < \log(I+3\times10^{-4}) < -4$.
}
\end{figure}

%\begin{figure}[ht!]
\begin{figure}
\center
 \includegraphics[width=5in]{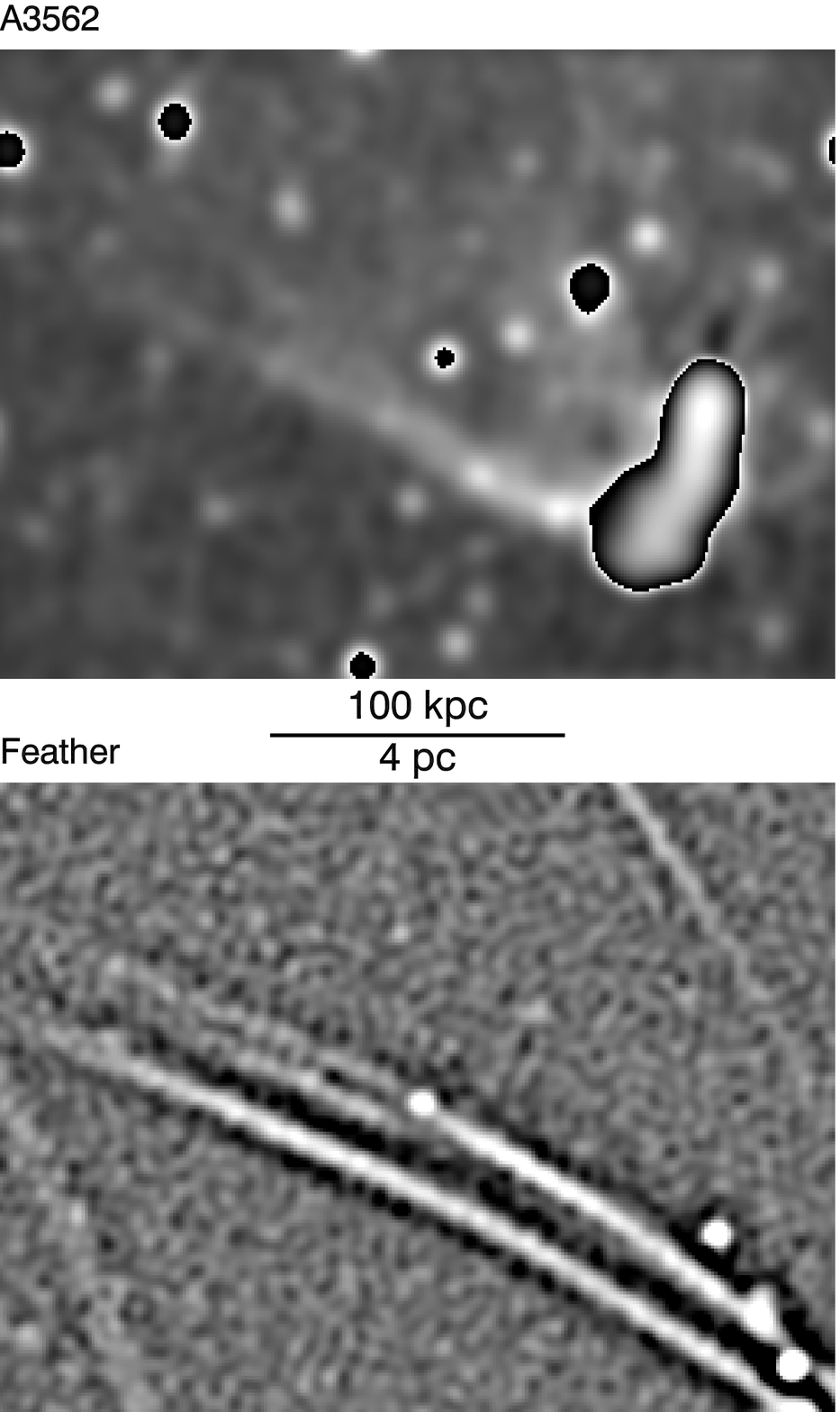}
 \caption{
{\it Top (a)}
A  MeerKAT image of the galaxy cluster A 3562 \citep{giaci22} convolved
with a Gaussian kernel with  FWHM $\sim10.3''$.
{\it Bottom (b)}
A portion of the Feather filament where the filament splits into two fainter components at
a compact source \citep{zadeh22b}.
The intensity I (Jy beam$^{-1}$) scales are listed as:
Feather: $-4.7 < \log(I+3\times10^{-4}) < -4.3$;
A3562: $-4 < \log(I) < -2 (I>7\times10^{-5})$.
}
\end{figure}

%\begin{figure}[ht!]
\begin{figure}
\center
\includegraphics[width=6.5in]{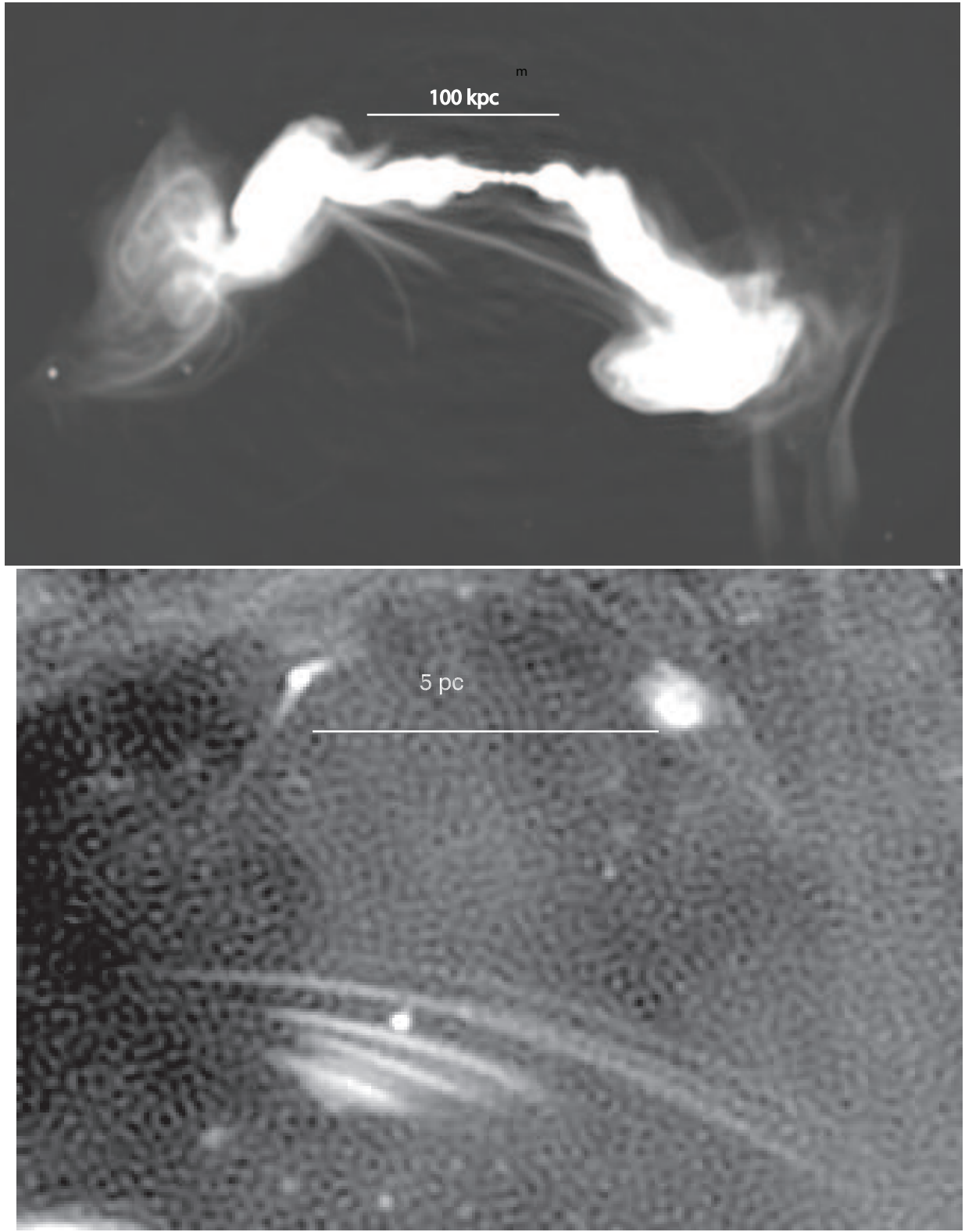}
 \caption{
{\it Top}
A MeerKAT image of ESO137-006  galaxy in Norma galaxy cluster
\citep{ramatsoku20}. The beam FWHM is  $\sim10''$. $7.62''\times7.29''$.  
{\it Bottom}
A portion of the Bent Harp  filament showing multiple parallel filaments \citep{zadeh22b}.
The intensity I (Jy beam$^{-1}$) scale for the 
Bent Harp is: $-3.8 < \log(I+3\times10^{-4}) < -3.$
}
\end{figure} 
\end{document}